\documentclass[12pt]{article}
\usepackage[utf8]{inputenc}
\usepackage[margin=1in]{geometry}
\usepackage{amsmath}
\usepackage{bm}
\usepackage{amsfonts}
\usepackage{amssymb}
\usepackage{graphicx}
\usepackage{cite}
\usepackage{xcolor}
\usepackage{hyperref}

\hypersetup{
	colorlinks=true,
	linkcolor=blue,
	filecolor=magenta,
	urlcolor=cyan,
	citecolor=green,
	pdfpagemode=FullScreen
}
\usepackage{setspace}
\date{}
\usepackage{authblk}

\title{\large\textbf{An analytical solution of Balitsky-Kovchegov equation using homotopy perturbation method} }

\author[1] {R. Saikia 
\footnote{Corresponding author:  \href{mailto:ranjans@tezu.ernet.in}{ ranjans@tezu.ernet.in}}}

\author[2] {P. Phukan \footnote{E-mail: \href{mailto:pragyan@morancollege.com}{ pragyan@morancollege.com}}}

\author[1] {J. K. Sarma \footnote{E-mail: \href{mailto:jks@tezu.ernet.in}{ jks@tezu.ernet.in}}}

\affil[1]{\small \emph{Department of Physics, Tezpur University, Tezpur, Assam, 784028, India}}

\affil[2]{\small \emph{Department of Physics, Moran College, Moranhat, Assam, 785670, India}}

\begin{document}
\maketitle

\begin{abstract}
  An approximate analytical solution of the Balitsky-Kovchegov (BK) equation using the homotopy perturbation method (HPM) is suggested in this work. We have carried out our work in perturbative QCD (pQCD) dipole picture of deep inelastic scattering (DIS). The BK equation in momentum space with some change of variables and truncation of the BFKL (Balitsky-Fadin-Kuraev-Lipatov) kernel can be reduced to the FKPP (Fisher-
  Kolmogorov-Petrovsky-Piscounov) equation [Munier and Peschanski $2003$]. The observed geometric scaling phenomena are similar to the travelling wave solution of the FKPP equation. We solved the BK equation using the HPM. The obtained solution in this work also suggests the travelling wave nature of the measured scattering amplitude $N(k, Y)$ plotted at various rapidities. The result obtained in this work can be helpful for different phenomenological studies in high-density QCD.  
	
	\vspace{1cm}
	
\end{abstract}
	
\thispagestyle{empty}

\clearpage

	\section{Introduction}
	     QCD evolution equations \cite{1,2,3,4,5,6,7} serve as an excellent tool in high energy physics phenomenology, which describes the evolution of the Parton Distribution Functions (PDFs) in a nuclear medium. According to perturbative QCD (pQCD), in light of high energy hadron collision experiments, it has been shown that as $x_{B}$ (Bjorken $x$) decreases, the growth of gluon density rapidly increases. This type of fast rapid growth of gluons is well described by the famous BFKL (Balitsky-Fadin-Kuraev-Lipatov) equation \cite{6,7}. This equation can be derived with pQCD by resuming leading logarithms of energies expressed in terms of $x_{B}$ such as $ln(1/x) >> ln(Q^2/\mu^2)$ ($Q$ is the photon virtuality, $\mu$ being the renormalization scale). In the solution of the BFKL equation, the scattering amplitude $N(k, Y)$ ($k$ being transverse momentum and $Y$ being the rapidity of evolved gluons) and hence the total cross-section exhibits an exponential growth with rapidity Y. As the energy increases more and more and hence correspondingly $x_{B}$ decreases, the gluon density more rapidly increases which need to be tamed down to hold the unitarity, as well as Froissart-Martin bound \cite{8}. According to Froissart-Martin bound, in QCD, the total cross-section of a process cannot grow faster than the logarithm of energy squared i.e., $ln^{2}s$ . Thus, at very high energies, the BFKL equation violates unitarity and Froissart bound; hence, its applicability is limited. We can not use this equation at arbitrarily high energies. \par 
	    To address the above problems faced by the BFKL equation, we need to look for solutions to the related problems. The solution is that at high energy and lower $x_{B}$, partons themselves start to recombine and get saturated. The first idea of parton-parton recombination was addressed in ref.\cite{9,10,11,12,13}. Some analytical solutions of nonlinear QCD evolution equations incorporating parton-parton recombination can be found in ref.\cite{14,15,16,17}. The parton-parton recombination will tame down the gluon density in the high gluon density region of the scattering process. BFKL equation, being linear, could not address this nonlinear effect of parton recombination and saturation and hence unable to explain underlying physics at high-density QCD. \par
	    
        It is imperative to understand the implicit physics in saturation regions of partons at low Bjorken $x_B$. In this region, nonlinear QCD evolution equations come into play which helps in understanding the physics in that region. Therefore, to describe the parton-parton recombination and saturation effect, the linear evolution equations have to be replaced by the nonlinear QCD evolution equations. The nonlinear evolution equation has important features dealing with the saturation effect. They contain damping terms that reflect the saturation effect arising out of parton-parton recombination. So, studying non-linear QCD evolution equations and their solutions is crucial for phenomenological studies. The JIMWLK (Jalilian-Marian-Iancu-McLerran-Weigert-Leonidov-Kovner) equation \cite{18,19,20,21} permits gluon saturation in a high gluon density region. JIMWLK equation addresses the nonlinear correction using the Wilson renormalization group approach. Nevertheless, it is complicated to solve the JIMWLK equation because of its complex nature. Instead, its mean-field approximation BK equation \cite{22,23,24,25} has been widely used in the context of saturation effect. Because of its simple nature, the BK equation can be solved numerically. However, it is tough to solve the BK equation using general methods. It is an integrodifferential equation in coordinate space that can be transformed into momentum space resulting in a partial differential equation. Solving the BK equation in momentum space can be useful for phenomenological studies in light of various high energy hadron scattering experiments. Analytical solutions to the BK equation proposed recently can be found with different approaches using some approximation in ref.\cite{26,27,28,29,30,31}. These analytical solutions shed light on the ability of the BK equation in explaining gluon saturation and its application in the high energy hadron scattering phenomena.\par
	    In this work, we have suggested an analytical solution to the BK equation using the homotopy perturbation method (HPM) \cite{32,33} in connection with the FKPP (Fisher-
	    Kolmogorov-Petrovsky-Piscounov) equation \cite{34,35}. The FKPP equation is a partial differential equation that belongs to the reaction-diffusion equation in statistical physics. The phenomena of geometric scaling at low $x_B$ observed at HERA can be related to the travelling wave solution of the FKPP equation \cite{36}. It is shown in the pioneering work \cite{26,27,28} that the BK equation in momentum space can be transformed into the FKPP equation with some variable transformation. It can be shown that the transition of the scattering amplitude into the saturation region is similar to the formation of the front of the travelling wave of the FKPP equation \cite{26}. The solution of the BK equation can be helpful for further phenomenological studies in light of present and future accelerator facilities.\par
	    We organize the paper as follows. Section \ref{sec:2} discusses the connection between the BK and the FKPP equations. In section \ref{sec:3}, an analytical solution of the BK equation is obtained from the HPM. Section \ref{sec:4} is devoted to discussion and summary.

\section{Connection between the BK and the FKPP equations}\label{sec:2}
The connection between the BK and the FKPP equations has been found in the pioneering work done by S. Munier and R. B. Peschanski \cite{26,27,28}. In this section, we discuss how to relate the BK equation with the FKPP equation following the work by them.\par 
BK equation is about energy dependence of scattering amplitude, it is often convenient to carry out work in the perturbative QCD (pQCD) dipole picture of deep inelastic scattering (DIS) \cite{37,38,39,40}. The main advantage of the dipole picture of DIS is the factorization of the scattering process into several steps. In the dipole picture, an incoming virtual photon after fluctuation changes to a quark-antiquark dipole. The quark-antiquark pair then scattered off the target proton and recombines to form some final state particles. In reference to the dipole picture of DIS, in the leading logarithm approximation of pQCD, the cross section in terms of the total rapidity ($Y$) and the virtuality of the photon ($Q$) factorizes to \cite{36}
\begin{equation}\label{eqn:1}
\sigma^{\gamma^{\ast} p}(Y,Q) = \int_{0}^{\infty}x_{01}dx_{01} \int_{0}^{1} dz |\psi(z, x_{01}Q)|^{2} N(Y, x_{01}),
\end{equation}
where $z$ being the longitudinal momentum fraction of the quark of the virtual photon, $\psi(z, x_{01}Q)$ is the photon wave function on a quark-antiquark dipole of its size $x_{01}$. $N(Y,x_{01})$ is the forward dipole-proton scattering amplitude.\par 
Within the large $N_c$ approximation at fixed coupling and for a homogeneous nuclear target, the measured scattering amplitude $\mathcal{N}(Y,k)$ at total rapidity $Y$ and transverse momentum $k$ obeys the BK equation in momentum space given by\cite{24}
\begin{equation}\label{eqn:2}
\partial_{Y}\mathcal{N} = \bar{\bm\alpha}\chi (-\partial_{L})\mathcal{N} - \bar{\bm\alpha}\mathcal{N}^{2} ,
\end{equation}
where $ \bar{\bm\alpha} = \frac{\alpha_s N_c}{\pi}$ and $\chi(\zeta) = 2\psi(1) - \psi(\zeta) - \psi(1-\zeta)$ is the BFKL kernel. $\zeta = - \partial_{L}$, where $L = \ln(\frac{k^2}{k^2_{0}})$, $k_0$ being some low momentum scale at fixed. The expansion of the BFKL kernel around $\zeta = \frac{1}{2}$ has been suggested in ref.\cite{26}, and with this expansion equation \eqref{eqn:2} reduces to the nonlinear partial differential equation given by,
\begin{equation}\label{eqn:3}
\partial_{Y}\mathcal{N} = \bar{\bm\alpha}\bar{\chi} (-\partial_{L})\mathcal{N} - \bar{\bm\alpha}\mathcal{N}^{2} ,
\end{equation}
where
\begin{equation}\label{eqn:4}
\bar{\chi}(-\partial_L) = \chi(\frac{1}{2}) + \frac{\chi''(\frac{1}{2})}{2}\left(\partial_L + \frac{1}{2}\right)^{2}.
\end{equation}
In reference to the above expansion and defining $\bar{\zeta} = 1 - \frac{1}{2} \sqrt{1 + 8\frac{\chi(\frac{1}{2})}{\chi''(\frac{1}{2})}}$, with the following change of variables\cite{26}
\begin{equation*}
t = \frac{\bar{\bm\alpha}\chi''(\frac{1}{2})}{2}(1 - \bar{\zeta})^2 Y, \hspace{5mm} x = (1 - \bar{\zeta})\left( L + \frac{\bar{\bm\alpha}\chi''(\frac{1}{2})}{2} Y \right),
\end{equation*}
\begin{equation*}
u(t,x) = \frac{2}{\chi''(\frac{1}{2})(1 - \bar{\zeta})^{2}} \mathcal{N}\left( \frac{2t}{\bar{\bm\alpha}\chi''(\frac{1}{2}) (1 - \bar{\zeta})^2}, \frac{x}{1 - \bar{\zeta}} - \frac{t}{(1 - \bar{\zeta})^{2}}\right),
\end{equation*}
the equation \eqref{eqn:3} turns into the FKPP equation\cite{34,35} for $u(x,t)$, can be expressed as\cite{26}
\begin{equation}\label{eqn:5}
\partial_t u(t,x) = \partial^2 _{x} u(t,x) + u(t,x) - u^2 (t,x).
\end{equation}
Thus, with some variable transformation, it is seen that the BK equation \eqref{eqn:3} can be transformed to the above equation \eqref{eqn:5}, which is the famous FKPP equation.

\section{Solution of BK equation with HPM}\label{sec:3}
Given the discussion of the connection between the FKPP and the BK equations discussed in the previous section, let us solve the BK equation for the scattering amplitude $\mathcal{N}(k, Y)$. The solution of the BK equation \eqref{eqn:3} in connection with the equation \eqref{eqn:5} for the scattering amplitude $\mathcal{N}(k, Y)$ using the HPM can be written as
 \begin{equation}\label{eqn:6}
 \mathcal{N}(k,Y) = \frac{\mathcal{N}_{0} e^Y }{1 - \mathcal{N}_{0} + \mathcal{N}_{0} e^Y},
 \end{equation}
 where $\mathcal{N}_{0}$ is the initial condition. Once the initial condition is known to us, the solution of the BK equation gives the scattering amplitude $\mathcal{N}(k, Y)$ at any given rapidity $Y > 0$. In this work, we will use the following initial condition given by K. Golec-Biernat and M. Wüsthoff (GBW), introduced first in ref.\cite{41},
 \begin{equation}\label{eqn:7}
 N^{GBW} (r, Y = 0) = 1 - exp \left[ -\left (\frac{r^2 Q^{2}_{s0}}{4} \right) \right].
 \end{equation}
 $Q^{2}_{s0}$ is the fit parameter, called the initial sasuration scale squared. The reason behind taking this initial condition is that as we are dealing with the BK equation in momentum space. This initial condition can be simply Fourier transformed into momentum space analytically. The momentum space result of the GBW initial condition can be written as 
 \begin{equation}\label{eqn:8}
 \mathcal{N}^{GBW}(k, Y=0) = \int \frac{d^{2}r}{2\pi r^{2}} e^{ik.r} N^{GBW}(r,Y=0) = \frac{1}{2} \Gamma\left(0, \frac{k^2}{Q^{2}_{s0}}\right).
 \end{equation}
 $\Gamma(0, {k^2}/{Q^{2}_{s0}})$ is the incomplete gamma function. At large values of $ {k^2}/{Q^{2}_{s0}}$, this behaves as 
 \begin{equation*}
 \Gamma\left(0, \frac{k^2}{Q^{2}_{s0}}\right) = exp \left(-\frac{k^2}{Q^{2}_{s0}}\right).
 \end{equation*}
 Therefore, 
 \begin{equation}\label{eqn:9}
 \mathcal{N}^{GBW}(k, Y=0) = \frac{1}{2}  exp \left(-\frac{k^2}{Q^{2}_{s0}}\right). 
 \end{equation}
 Substitution of the above equation in equation \eqref{eqn:6} for the initial condition $\mathcal{N}_0$, we obtain the scattering amplitude $\mathcal{N} (k, Y)$ with GBW as the initial condition 
 \begin{equation}\label{eqn:10}
 \mathcal{N}(k, Y) = \frac{e^{Y-k^2/Q^{2}_{s0}}}{1 - e^{-k^2/Q^{2}_{s0}} + e^{Y-k^2/Q^{2}_{s0}}}.
 \end{equation}
 This is the approximate analytical solution of the BK equation \eqref{eqn:3}. The evolution of the scattering amplitude at different rapidities can be seen in the  Figure 1.
 \begin{figure}[h]
 	\centering
 	\includegraphics[scale=0.87]{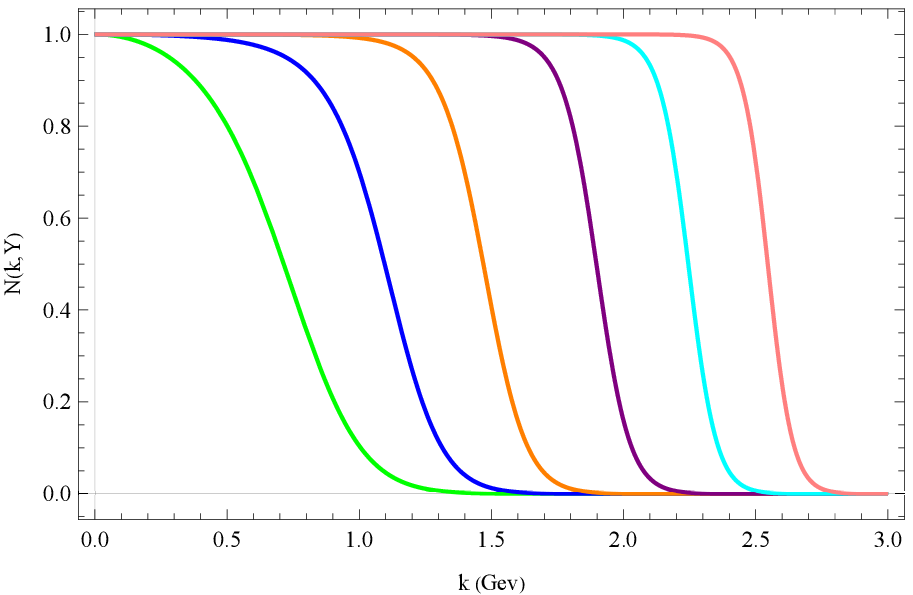}
 	\caption{The solution of the BK equation in momentum space, $\mathcal{N}(k)$, at various rapidities $Y=\textcolor{green}{2}$, $Y=\textcolor{blue}{5}$, $Y=\textcolor{orange}{9}$, $Y=\textcolor{purple}{15}$, $Y=\textcolor{cyan}{21}$ and $Y=\textcolor{pink}{27}$. }
 	\label{fig:1}

\end{figure}
 
\section{Discussion and Summary}\label{sec:4}
This work has suggested an approximate analytical solution of the BK equation using the HPM. The connection between the geometric scaling phenomena of the solution of the BK equation and the travelling wave solution of the FKPP equation, as suggested by S. Munier and R. Peschanski in their pioneering work, has guided the scientific community working in the field of saturation physics. In this letter, we have started our discussion with the connection between the BK and FKPP equation. We carried out work in the pQCD dipole picture of DIS in which the measured scattering amplitude $\mathcal{N}(k, Y)$ obeys the BK equation in momentum space. The momentum space frame is often considered the natural space frame to work in the context of at least a travelling wave solution and the geometric scaling. Afterward, with some change of variables and a slight approximation in the BK equation, we ended with the approximated analytical solution of the BK equation in the momentum space. We have plotted the obtained solution, equation \eqref{eqn:10}, at different rapidities in Figure \ref{fig:1} to check the travelling wave nature of the solution. Indeed, one can see the solution's travelling wave nature. It indicates that at high energy, the scattering amplitude behaves as a wave travelling from the region $\mathcal{N} = 1$ to $\mathcal{N}=0$ as $k$ increases without being changed in the profile. This is indeed a vital physical result of this travelling wave approach.\par
The solution obtained in this work can be helpful in further phenomenological studies on high-density QCD and saturation regions. However, it is going to be interesting to observe whether this type of travelling wave solution and geometric scaling exist or not at very high energies when EIC (Electron-Ion Collider)\cite{42} and other future projects run operation. Nevertheless, the BK equation with truncation of the BFKL kernel successfully explains the observed geometric scaling and the travelling wave nature of its solution at current accelerator facilities. We must rely on future accelerator facilities for precise measurements of observed phenomena and their confirmation.

\begin{center}
\rule{8cm}{0.3mm}
\end{center}

\end{document}